# Understanding Magic Numbers in Neutron-Rich Nuclei by Tensor Blocking Mechanism

I. Tanihata[a,b], H. Toki[b], S. Terashima[a], and H.-J. Ong[b]

[a] IRCNPC and School of Physics, Beihang University, Beijing 100191 , P. R. China

[b] RCNP, Osaka University, Osaka 567-0047, Japan

## ABSTRACT

A new paradigm for nuclear structure that includes blocking effects of tensor interactions is proposed. All of the recently discovered magic numbers ($N$=6, 14, 16, 32 and 34) in neutron-rich nuclei can be explained by the blocking effects. A large amount of binding energy is gained by high-momentum correlated pairs of nucleons produced by the tensor interaction. Such tensor correlations strongly depend on the configuration space available for exciting nucleons to 2p-2h states. When additional neutrons occupy a new orbital, the previously available configuration may be lost, resulting in a sudden loss of binding energy otherwise gained by the 2p-2h excitations. Such tensor blocking effects enlarge the energy gaps at all observed new magic numbers. Tensor blocking also explains consistently the observed peculiar configurations of neutron-rich nuclei at the borders of shells.

## 1. Introduction

The Pauli principle is essential in the formation of quantum-mechanical many-body systems. In nuclear physics it is the basis for forming shell structures and magic numbers. The nuclear shell model is based on single-particle orbitals in one-body potentials and the filling of the orbitals satisfying the Pauli principle. The structure of a nucleus is understood by the occupation of nucleons in such orbitals. The magic number, which is an important aspect of the shell model, are understood to be fluctuations of the energy gaps between orbitals. In the present paper, we discuss the effects of tensor correlations on the spacing of orbitals and magic numbers.

The structures of neutron-rich nuclei are being quickly studied after beams of radioactive nuclei became available [1]. Many new properties of nuclei have been observed. One of the striking findings are new magic numbers $N$=6, 14, 16, 32, 34, and the disappearance of the standard magic numbers $N$=8, 20 in neutron rich nuclei [2,3,4,5,6,7,8]. In addition to such matters related to the magic numbers, peculiar behaviors of particle configurations are observed in $^{11}$Li, $^{11}$Be and $^{12}$Be nuclei. The most classical case is the inversion of the positive (1/2+) and negative (1/2-) parity states in $^{11}$Be. In $^{11}$Li the ground state (two-neutron halo state) is observed to have $(2s_{1/2})^2$ and $(1p_{1/2})^2$ configurations mixed almost equally [9]. These peculiarities indicate a lowering of the $2s_{1/2}$ orbitals along $N$=8 for nuclei lighter than the oxygen. On the other hand, a recent study shows a dominance of $(1d_{5/2})^2$ and much weaker contributions from $(2s_{1/2})^2$ and $(1p_{3/2})^2$ in $^{12}$Be ground state [10,11,12]. Each of the phenomena described above has been studied extensively and some suggested models can reproduce each phenomenon individually. However, no unified principle for understanding these phenomena has been formulated.

It is well known that the strong tensor interaction is caused by pion exchange between nucleons due to the pseudo-scalar nature of the pion. The saturation property of nuclear matter is caused by the blocking effect of the tensor interaction, which was demonstrated by Bethe using the Bruckner–Hartree–Fock theory [13]. The tensor interaction excites a proton and neutron pair in the Fermi sea to a pair with intermediate or large relative momentum (2p-2h excitation) outside the Fermi sea, and provides a large binding effect on nuclear matter. With the increase of density, the Fermi momentum increases and some of the intermediate momentum states used by the tensor interaction start to be blocked by the nucleons in the Fermi sea. This blocking effect decreases the binding energy per nucleon, and eventually provides the saturation property for the nuclear matter. It has two important consequences on nuclear spectroscopy. The first is the appearance of a p–n pair with intermediate or large relative momenta (tensor states) in the excitation spectra of finite nuclei [14,15,16]. The other is an energy increase of shell-model states by blocking the tensor states by filled nucleons, where the tensor states are those used by 2p-2h excitations by the tensor interaction.

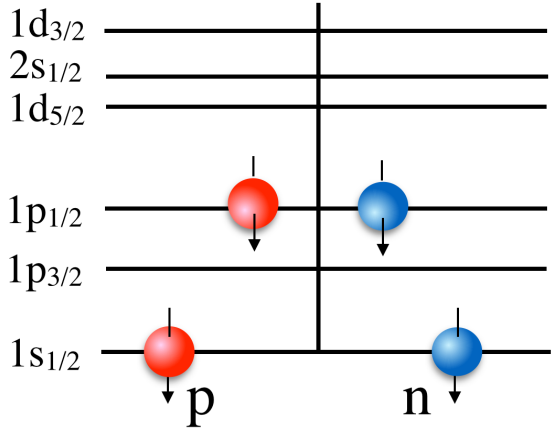

Fig. 1 The most important 2p-2h excitation by the tensor interaction in $^4$He. This configuration alone, $(1p_{1/2})^2(1s_{1/2})^{-2}$, provides 8.4 MeV of potential energy in a $^4$He nucleus.

In finite nuclear system, the lightest nucleus is the deuteron, made of a proton and a neutron. In addition to S wave its wave function has a large D-wave component caused by the tensor interaction. The tensor interaction provides considerable potential energy and the deuteron is bound at a binding energy of 2.2 MeV [17]. The $^4$He nucleus has a large binding energy of 28.3 MeV. A large fraction of the binding energy is also caused by the tensor interaction [18]. The total potential energy gained by the tensor interaction is 54.6 MeV using the AV8′ potential in the tensor optimized shell model [19]. Two-particle two-hole excitations by the tensor interaction include excitations to high $l$ orbitals. However, the tensor optimized shell model (TOSM) calculations showed that the excitations to $l=1$ orbitals already provide a large potential energy, and among them the greatest contribution comes from a 2p-2h excitation from the $(1s_{1/2})^2$ state to the $(1p_{1/2})^2$ state providing 8.4MeV of potential energy as shown in Fig. 1 [19].

When an orbital is occupied by nucleons, tensor blocking occurs because the 2p-2h excitation to the tensor state is blocked by the occupying nucleon. When the $1p_{1/2}$ orbital is occupied in $^{16}$O, for example, the $(1s_{1/2})^2$ to $(1p_{1/2})^2$ excitation is blocked and the nucleus loses the potential energy gained otherwise. In $^{16}$O (a symmetric nucleus), however, the added proton and neutron pairs in the $1p_{1/2}$ shell open new excitations to the sd shell and gain potential energy. When the mass number increases along the line of stability, tensor blocking and tensor opening occur simultaneously, giving rise to the saturation property of the nuclear binding. This is exactly the same mechanism that produces the saturation property of nuclear matter [13]. It should be noted, however, that the tensor blocking occurs even if only a neutron orbital is filled. When only the neutron number increases, tensor blocking occurs but tensor opening does not occur. It therefore makes a difference in the effects of the tensor interactions between $Z\sim N$ nuclei and neutron-rich nuclei.

The tensor blocking effect in nuclei has been demonstrated by Myo et al. in $^{11}$Li [20]. To form $^{11}$Li, two-neutrons are added to $^9$Li in which the 1p$_{3/2}$ orbitals are occupied and thus the two neutrons have to be added to higher orbitals. Although the normal shell model predicts 1p$_{1/2}$ to be the next lowest orbital, tensor blocking occurs if neutrons occupy the 1p$_{1/2}$ orbital and thus the occupation of the 1p$_{1/2}$ orbital is disfavored. The contribution of 2s$_{1/2}$, the next expected orbital, then become important because no tensor blocking occurs even if the orbital is filled.

To calculate the tensor effect fully in finite nuclei, we have to treat intermediate- and large-momentum pair states in the wave function. This work has been performed in the few body framework [18], the GFMC method of the Argonne group [21], and the TOSM and TOAMD methods [20,22]. Experimentally, high-momentum neutrons and tensor correlated high-momentum p–n pairs have been reported [14,15]. There is another interesting effect of the tensor interaction, which provides a spin–orbit effect locally for spin non-saturated nuclei in the shell model framework through the exchange diagram of the tensor interaction called the monopole term [23]. The tensor interactions between the $j_p$ and $j_n$ states are calculated through the exchange diagram, hence, we call this tensor mechanism the tensor Fock mechanism. This mechanism works in an opposite way from the blocking effect discussed in this paper. The tensor Fock effect is completely zero for spin closed shell nuclei (Z/N=2, 8, 20, 40,…) of either protons or neutrons. In contrast, the tensor blocking mechanism has a large effect on the binding energy and the spin–orbit splitting in any nuclear system [24].

## 2. The tensor blocking shell model

To include the effect of the tensor interaction through high-momentum correlated nucleon-pairs, we introduce a simple model based on the TOSM [25,26], which includes 2p-2h excitations explicitly in the wave function,

$$\Psi = \psi_{Sh} + \psi_{2p\text{-}2h}, \qquad (1)$$

where $\psi_{Sh}$ is the usual shell model wave function and $\psi_{2p\text{-}2h}$ expresses the 2p-2h states excited by the tensor interaction and includes high-momentum proton–neutron pairs. The Hamiltonian includes both the central and the tensor interactions written as $H = H_C + H_T$, where $H_C$ contains the kinetic energy, the central interactions, the spin–orbit interactions, and other non-tensor interactions and $H_T$ denotes the tensor interaction. TOSM includes 2p-2h excitations explicitly and considers bare interactions between nucleons. The energy of a nucleus is calculated by

$$\langle\Psi|H_c + H_T|\Psi\rangle = \langle\psi_{Sh}|H_c|\psi_{Sh}\rangle + \langle\psi_{Sh}|H_T|\psi_{Sh}\rangle + 2\langle\psi_{Sh}|H_T|\psi_{2p\text{-}2h}\rangle + \langle\psi_{2p\text{-}2h}|H_c+H_T|\psi_{2p\text{-}2h}\rangle. \qquad (2)$$

The second term on the right-hand side is the monopole term arising from exchange interactions (Fock term) of the tensor interaction, and has an important contribution to the change of $l \cdot s$ splitting in neutron-rich nuclei [23]. The last two terms are contributions from the 2p-2h excitations in TOSM, out of which the third term (transition term) provides a large attraction. For example, for the $^4$He nucleus [19] 2p-2h excitations provide 55 MeV of attraction from the tensor interaction, which is comparable with the potential energy of 56 MeV from the central interactions in the first term in eq. (2). $\psi_{2p\text{-}2h}$ includes excitations to higher $l$ orbitals. However more than 10 MeV comes from 2p-2h excitations to $l=1$ alone in $^4$He. Higher excitations provide remaining potential energy of ~45 MeV. However, as seen in TOSM calculations, the inclusion of higher $l$ excitations require considerable computer power and human resources, and it is time consuming in practice to complete the calculation for nuclei above B isotopes [25, 26].

In light neutron-rich nuclei, neutrons occupy the next major shell but $\Delta l \geq 2$ shells are always open. The blocking of 2p-2h excitations by the occupation of nucleons, thus, does not occur for $\Delta l \geq 2$ excitations. The blocking occurs only for $\Delta l=1$ 2p-2h excitations, and therefore we take only these states in the model space with the wave function written as $\psi_{Sh} + \psi_{2p\text{-}2h}^{(\Delta l=1)}$. In usual shell models without 2p-2h excitations, the binding energy due to the tensor interaction is effectively included in $H_C$ by the model setting. In the present tensor blocking shell model, we explicitly discuss the tensor interaction by 2p-2h excitations to $\Delta l=1$ orbitals, which could be blocked in the shell model states. Therefore, the other $\Delta l \geq 2$ 2p-2h excitations are considered to provide only a smooth change in the potential energy. In principle we can calculated their contributions using the Feshbach projection method to modify the bare interaction to an effective interaction [27].

In the following discussion, the Woods–Saxon potential with standard spin–orbit coupling is used as the effective potential to obtain the starting single-particle orbitals. The starting single-particle orbitals at $A/Z=3$, as an example, are shown in Fig. 2 in which the potential parameters are taken from Bohr–Mottelson [28]. We consider that most of the potential energy obtained by the tensor interaction is effectively included already in this Woods–Saxon potential for ordinary nuclei. It is noted that the order and spacing of the single-particle orbitals change depending on the binding energy. The low angular momentum orbitals gain binding energy under weakly bound conditions, which is a halo effect.

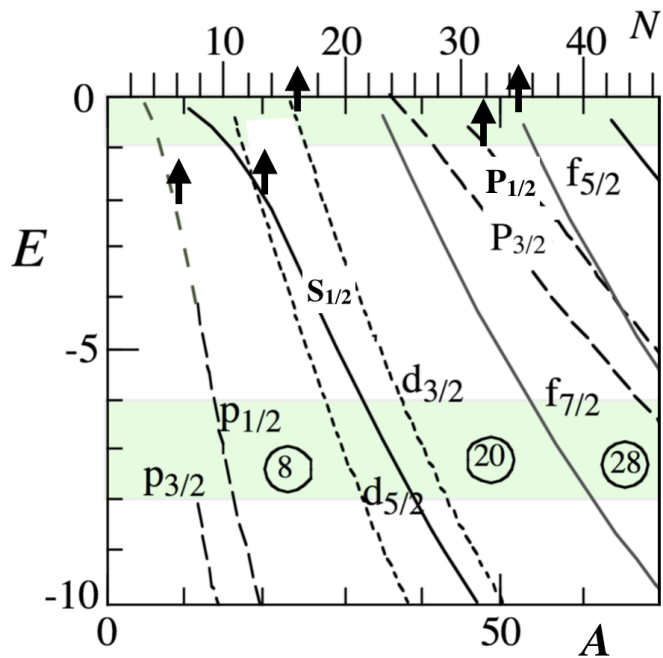

Fig. 2 Single particle orbitals in a Woods-Saxon potential. The potential parameters are taken from Bohr-Mottelson [21]. Note that the order of single particle orbitals changes depending on the binding energy. The numbers in circles inidicate traditional magic numbers. Arrows indicate the position and the direction of effects of tensor blocking.

The tensor blocking shell model treats separately the sudden change of the binding energy of a nucleus due to tensor blocking. Important 2p-2h excitations and their potential energies are selected from the results of TOSM [19,20,22,25,26]. In the valence shell, the largest contribution comes from the excitation of $S=1$ p–n pairs from the $(n,l,j)^2$ configuration to $(n+1,l\pm 1,j)^2$. Blocking of such excitations reduces the binding energy considerably according to TOSM calculations. By this effect, the filling of weakly bound orbitals is drastically affected. However, it should be noted that the order and spacing of deeply bound single-particle orbitals are not affected significantly because the energy change due to the blocking is smaller than the total depth of the potential. The order of the original single particle orbitals is kept intact in the following discussions.

Part of the $\boldsymbol{l\cdot s}$ splitting is known to originate from the tensor interaction [24]. The present blocking effects affect the splitting between $j_>$ and $j_<$ orbitals and would contribute to the $\boldsymbol{l\cdot s}$ splitting. We do not know presently, however, how to remove the effect of the tensor interaction in the shell model $l \cdot s$ splitting. In the present model, therefore, we used the usual $l \cdot s$ splitting without modification. In the following we semi-quantitatively discuss the ground state properties of light neutron-rich nuclei under the tensor-blocking shell model.

### 3. Loss of traditional magic numbers and appearance of new magic numbers

We discuss first the appearance and disappearance of magic numbers in neutron rich nuclei. To this end we include a table for Woods-Saxon single particle energies and tensor blocking energies in Table 1. The Woods-Saxon energies are obtained by solving single particle states of nuclei using a single-particle potential with the Bohr-Mottelson parameters. This parameter set is suitable to discuss the tensor blocking effect, since the parameters are obtained from stable and near stable nuclei. As for the tensor blocking (TB) energies, we take $2\Delta E_{TB}$= 5 MeV for the $^{10}$He nucleus and add $A^{-1/3}$ dependence on the TB energy. As for one neutron occupying a tensor state as the single particle energy of $^{10}$Be, the tensor blocking energy is $\Delta E_{TB}$= 2.5 MeV. The TB energy is exactly the same as the energy gained by the 2p-2h excitation except the sign when the corresponding orbital is empty that can be calculated by the TOSM.

Table 1. Single particle energies and effects of tensor blocking.

| Nucleus | Orbit | Woods-Saxon energy [MeV] | Tensor blocking (TB) [MeV] |
|---|---|---|---|
| $^{10}$He | $p_{1/2}$ | $E(p_{1/2}) > 0$ (=-0.49) | 5.0 → unbound |
| $^{22}$O | $s_{1/2}$ | $E(s_{1/2}) - E(d_{5/2}) = 2.8$ | 1.9 → 4.7 MeV |
| $^{24}$O | $d_{3/2}$ | $E(d_{3/2}) - E(s_{1/2}) = 3.7$ | 1.9 → 5.6 MeV |
| $^{52}$Ca | $p_{1/2}$ | $E(p_{1/2}) - E(p_{3/2}) = 2.3$ | 1.4 → 3.7 MeV |
| $^{54}$Ca | $f_{5/2}$ | $E(f_{5/2}) - E(p_{1/2}) = 0.8$ | 1.4 → 2.2 MeV |
| $^{10}$Be | $p_{1/2}$ | $E(s_{1/2}) - E(p_{1/2}) > 3.8$ | 2.5 → 1.3 MeV |
| $^{12}$Be | $p_{1/2}$ | $E(d_{5/2}) - E(p_{1/2}) > 4.2$ | 2.4 → 1.8 MeV |
| $^{30}$Ne | $d_{3/2}$ | $E(f_{7/2}) - E(d_{3/2}) = 3.9$ | 1.7 → 2.2 MeV |

For the TB energy, we have studied all the configurations published for the He isotopes, Li isotopes and $^{8}$Be in the TOSM framework [19,22,25,26]. In $^{4}$He, the tensor attraction due to $(1p_{1/2})^2(1s_{1/2})^{-2}$ excitation is 8.5 MeV and about a few MeV energy is lost due to the increase of kinetic energy [19]. In $^{6}$He, the first and second $0^+$ states have the energy difference due to the tensor interaction by 12 MeV due to the fact that the first $0^+$ is made of $(1p_{3/2})^2$ and the second $0^+$ is made of $(1p_{1/2})^2$ predominantly. In $^{8}$He, the energy difference between two $0^+$ states is about 9 MeV. This difference should be caused by the tensor blocking, since one of the two states does not occupy the $1p_{1/2}$ state [22]. In $^{6}$Li, the difference of tensor matrix elements between the first and second $1^+$ states is 5.4 MeV and we see several states having neutrons in the $1p_{1/2}$ state have less binding energies [25,26]. Seeing all these numbers, we have taken $2\Delta E_{TB}$= 5 MeV for the $^{10}$He nucleus as the tensor blocking energy. As for the $A$ dependence, we consider the effect of the associated wave functions increase their extensions by $A^{1/3}$ as the nuclear radius. The tensor matrix elements require short-range part of wave functions and we should therefore take the matrix elements behave as scaling with $A^{-1/3}$ [30].

An example of an abrupt change of binding is seen in He isotopes. Neutrons occupy the $p_{3/2}$ orbital mostly until $^{8}$He, and further addition of neutrons is expected to fill the bound $1p_{1/2}$ orbital in the Woods–Saxon potential. However, the filling of the $1p_{1/2}$ orbital blocks the 2p-2h excitation from the $1s_{1/2}$ orbital to the $1p_{1/2}$ orbital and thus a loss of binding energy occurs. The loss of energy is larger than the binding energy of the $1p_{1/2}$ orbital and thus creates a large energy gap above $N=6$, making $N=6$ magic and $^{10}$He unbound. In the case of $^{10}$He with the Woods-Saxon potential, $1p_{1/2}$ is unbound as seen in Table 1 and definitely the tensor blocking effect pushes further the nucleus unbound. The Bohr-Mottelson parameters of the Woods-Saxon potential provide a shallow binding of $1s_{1/2}$ state for He isotopes. For example $1s_{1/2}$ orbital is bound only by 17 MeV for $^{4}$He. To simulate the reality we have modified the potential by 10% deeper for He isotopes. Then $1p_{1/2}$ is bound by 0.49 MeV in $^{10}$He, which is shown in bracket in Table 1. Even this is the case, $^{10}$He is unbound by the tensor blocking, whose energy is $2\Delta E_{TB}=5$ MeV.

A similar blocking effect occurs in O isotopes. Six neutrons in $^{22}$O mostly occupy the $1d_{5/2}$ orbital. It should be noted that $1d_{5/2}$ orbital is not used as a tensor state and thus no tensor blocking occurs. The open $2s_{1/2}$ orbital is used for 2p-2h excitation from $1p_{1/2}$ to $2s_{1/2}$ orbitals. Another 2p-2h excitation from $1p_{3/2}$ to $1d_{3/2}$ is also open. Adding neutrons to either of the orbital blocks the tensor interaction and widens the energy gap. The numbers in Table 1 show that the energy gap between $2s_{1/2}$ and $1d_{5/2}$ was 2.8 MeV and the TB energy on $2s_{1/2}$ is $\Delta E_{TB}=1.9$ MeV makes the gap energy of 4.7 MeV. The energy gap thus produced makes $N=14$ a magic number. After filling two neutrons in $2s_{1/2}$ after the gap, the addition of neutrons to the $1d_{3/2}$ orbital again blocks the tensor interaction and thus another energy gap occurs at $N=16$, creating another magic number. We get from the single particle spectrum the energy difference between $2s_{1/2}$ and $1d_{3/2}$ is 3.7 MeV in $^{24}$O. The TB energy is added to the $1d_{3/2}$ state and the energy gap becomes 5.6 MeV. The addition of protons (F isotopes), instead, opens new configurations for the 2p-2h excitations, the excitation of a sd-shell proton–neutron pair into the fp-shell. It therefore suddenly increases the binding of nuclei compared with neutron-rich O isotopes, and thus makes the dripline of F isotopes extend much more than that of O isotopes.

The same mechanism works for Ca isotopes, as seen in the expected shell orbitals in Fig. 2. In Ca isotopes, the open shells $2p_{3/2}$, $2p_{1/2}$ and $1f_{5/2}$ are used for 2p-2h excitations from the sd shell. Therefore, tensor blocking occurs in those orbitals and larger energy gaps are created for neutron-rich isotopes of Ca at $N=28$. $^{48}$Ca, the first asymmetric doubly closed shell nuclei in the neutron-rich region, is also supported by 2p-2h excitations by tensor interactions. Tensor blocking does not occur at the $1f_{7/2}$ orbital but occurs at the $2p_{3/2}$ orbital and enlarges the gap at $N=28$. Blocking occurs for $^{10}$He and $^{28}$O because they have $LS$ closed shells. However, blocking does not occur in the $jj$ closed $^{48}$Ca. We discuss new magic numbers for Ca isotopes. Shown in Table 1 is the energy difference of $2p_{1/2}$ and $2p_{3/2}$ states in $^{52}$Ca, which is 2.3 MeV. The tensor blocking energy on $2p1/2$ state is $\Delta E_{TB}=1.4$ MeV. The net magic energy becomes 3.7 MeV for the $N=32$ magic energy. In $^{54}$Ca, neutrons occupy until $2p_{1/2}$ state, and the energy difference of $1f_{5/2}$ and $2p_{1/2}$ is 0.8 MeV as shown in Table 1. The TB energy of $\Delta E_{TB}=1.4$ MeV brings the gap energy of 2.2 MeV. This energy difference of 2.2 MeV may be small to be called as to produce the magic effect, but the tensor blocking mechanism certainly helps to provide the magic effect at $N=34$.

### 4. Peculiar configurations in $^{11}$Li, $^{11}$Be, and $^{12}$Be

Peculiar behaviors of ground state spin and configurations are known along the neutron-rich $N=8$ region. The ground state of $^{11}$Li is considered to have an equal mixing of $(2s_{1/2})^2$ and $(1p_{1/2})^2$ orbitals. $^{11}$Be has an abnormal ground state spin-parity of $1/2^+$. On the other hand, the ground state configuration of $^{12}$Be is dominated by $(1d_{5/2})^2$ with a much weaker mixing of $(2s_{1/2})^2$ and $(1p_{1/2})^2$. Sudden changes in the relative position of orbitals have to be explained to understand such behavior in the shell model. Here we show that these changes are consistent with the tensor blocking mechanism.

Tensor blocking was first introduced by Myo et al. [20] to explain the equal mixing of s and p waves in the $^{11}$Li nucleus based on the TOSM calculations as already been discussed above. The ground sate of $^{11}$Be has a spin-parity of $1/2^+$, suggesting a lowering of the $2s_{1/2}$ orbital below the $1p_{1/2}$ orbital. Because of the very weak binding ($S_n=0.50$ MeV), the $2s_{1/2}$ neutron orbital is lower than the $1d_{5/2}$ orbital and close to the $1p_{1/2}$ orbital in the usual mean field potential (see Fig. 2) but still not lower than $1p_{1/2}$. Although deformation or a two-alpha character of Be isotope can be introduced to explain the additional lowering of $1/2^+$ [29], we show an additional reason. In $^{11}$Be, tensor blocking occurs for the $1p_{1/2}$ orbital but does not occur for the $2s_{1/2}$ orbital and thus contributes to the further relative lowering of the $1/2^+$ state in $^{11}$Be. In Table 1, we show the energy difference of $1s_{1/2}$ and $1p_{1/2}$ is more than 3.8 MeV, since $1s_{1/2}$ is not bound. Since experimentally $1s_{1/2}$ is bound at the threshold energy, we assume the energy difference is 3.8 MeV and the additional tensor blocking energy of $\Delta E_{TB}$=2.5 MeV brings this difference to 1.3 MeV.

The ground state of $^{12}$Be has a large spectroscopic factor of $(1d_{5/2})^2$, about 60%, compared with those of $(1p_{1/2})^2$ and $(2s_{1/2})^2$ that each contribute about 20% of the intensity [10]. For $^{12}$Be, the separation energies of neutrons are larger ($S_n$=3.17 MeV, $S_{2n}$=3.67 MeV), and thus the $1d_{5/2}$ orbital is lower than $2s_{1/2}$ in the normal Woods–Saxon potential (Fig. 2). The neutron occupation in the $1p_{1/2}$ orbital makes tensor blocking so that this configuration moves to a higher energy. Neutron occupation in $1d_{5/2}$ does not cause blocking and thus provides the lowest energy. The relation between $1p_{1/2}$ and $2s_{1/2}$ orbital is similar to that of $^{11}$Li giving a smaller contribution to the ground state of $^{12}$Be. For the case of the Woods-Saxon potential for $^{12}$Be, $1d_{5/2}$ state is unbound as seen in Table 1. Supposing the $1d_{5/2}$ is bound near the threshold, the tensor blocking energy on $1p_{1/2}$ makes the energy difference at 1.8 MeV. The magic gap at $N$=8 seems to be removed by the tensor blocking mechanism.

## 5. Break down of the magic numbers in neutron-rich nuclei

As discussed above, the tensor blocking mechanism helps to mix the 1p shell and 2s1d shell when the numbers of protons and neutrons are very asymmetric. When the proton number in the p-shell is none or small, before filling the $1p_{1/2}$ orbital significantly, a neutron in the $1p_{1/2}$ orbital exhibits tensor blocking, while the $1d_{5/2}$ and $2s_{1/2}$ orbitals do not cause tensor blocking. The tensor blocking helps to narrow or even close the shell gap originally expected by the shell model at $N$=8.

A similar phenomenon is expected for $N$=20 neutron-rich nuclei just above O. While the $1d_{3/2}$ neutron orbital exhibits blocking, the upper $1f_{7/2}$ and $2p_{3/2}$ orbitals do not cause blocking. The tensor blocking helps mix the fp shell orbitals into the sd shell and thus should contribute to making an island of inversion. It is suggestive to see that the limit of the island is at around Si, where proton starts to fill $2s_{1/2}$ shells and open new 2p-2h excitations. We provide shel-model single-particle energies and the associated tensor blocking energies in Table 1 for further discussions on the tensor shell model. As an example we show the case of $^{30}$Ne, where the gap energy at $N$=20 is 3.9 MeV. Adding the tensor blocking energy of $\Delta E_{TB}$=1.7 MeV, the difference becomes 2.2 MeV. It may be considered that the proximity of $1d_{3/2}$ and fp orbitals is the cause of the deformation in the island of inversion.

## 6. Summary and conclusion

In summary, we examined recently discovered magic numbers and peculiar changes of configurations in neutron-rich nuclei under a new paradigm of 2p-2h excitations (or high-momentum correlated p–n pairs) based on an explicit treatment of the tensor interaction in nuclei. All the new magic numbers $N$ = 6, 14, 16, 32 and 34 could be understood as being due to blocking of the tensor interaction. The blocking generates abrupt changes of binding energy and forms an additional energy gap at the neutron-rich region. It is understood that this blocking mechanism effectively works only for very asymmetric nuclei. In the present study, it was found that high-momentum nucleons and tensor blocking play essential roles in the structure of asymmetric nuclei. For the estimation of the shell structure we used the Woods-Saxon single particle spectrum with the Bohr-Mottelson parameters. We my have to replace the nuclear structure mode without the tensor blocking mechanism as the one of the relativistic mean field theory. Theoretical developments that include an explicit treatment of the tensor interaction through two-particle two-hole excitations of high-momentum proton–neutron pairs are anticipated for a more quantitative understanding.

*Acknowledgements*

We are grateful to Prof. Myo for fruitful discussions and for the use of the TOSM program in the present project. The experiment was partly supported by a grant-in-aid program of the Japanese government under the Contracts numbers 20244030, 23224008. and 14J03935. The support of the P.R. China government and Beihang University under the Thousand Talent program is gratefully acknowledged. This work is partially supported by the National Science Foundation of China under contracts numbers. 11235002, 11375023, 11475014, and 11575018 and by the National Key R&D program of China (2016YFA0400504).

**References**

[1] I. Tanihata, Nucl. Instr. and Meth. in Phys. Res. B **266**, 4067 (2008) and references therein.

[2] "Special issue on research opportunities with accelerated beams of radioactive ions" Nucl. Phys. A **693** (2001), edited by I. Tanihata.

[3] A. Ozawa et al., Phys. Rev. Letters **84**, 5493 (2000).

[4] C. Samanta and S. Adhikari, Phys. Rev. C **65**, 037301 (2002).

[5] D. Steppenbeck et al., Nature 12522 (2013).

[6] D.-C. Dinca et al., Phys. Rev. C **71**, 041302(R) (2005).

[7] H. Scheit, J. Phys. Conf. Ser. **312**, 092010 (2011).

[8] A. O. Macchiavelli et al., Phys. Rev. C **96**, 054302 (2017).

[9] I. Tanihata, H. Savajols, and R. Kanungo, Progr. in Part. and Nucl. Phys **68,** 215 (2013)

[10] J. Chen et al., Phys. Letters B **781**, 412 (2018). J. Chen et al., Phys. Rev. C **98**, 014616 (2018).

[11] A. Navin et al., Phys. Rev. Letters **85**, 266 (2000).

[12] S. D. Pain et al., Phys. Rev. Letters **96**, 032502 (2006).

[13] H. A. Bethe, Ann. Rev. Nucl. and Parti. Sci. **21**, 93 (1971).

[14] H. J. Ong et al., Phys. Letters B **725**, 277 (2013).

[15] S. Terashima et al., Phys. Rev. Letters **121**, 242501 (2018).

[16] R. Subedi et al., Science **320**, 1476 (2008).

[17] K. Ikeda et al., Clusters in Nuclei 165 (2011).

[18] H. Kamada, Phys. Rev. C **64**, 044001 (2001).

[19] T. Myo, H. Toki and K. Ikeda, Progr. Theor. Phys. **121**, 511 (2009).

[20] T. Myo, K. Kato, H. Toki and K. Ikeda, Phys. Rev. C **76**, 024305 (2007).

[21] R. B. Wiringa, S. C. Pieper, J. Carlson and V. R. Pandharipande, Phys. Rev. C **62**, 014001 (2000). R. B. Wiringa, Progr. Thor. Phys. Suppl. **146**, 403 (2002).

[22] T. Myo et al., Progr. Theor. and Exp. Phys. 073D02 (2015).

[23] T. Otsuka et al., Phys. Rev. Letters **95**, 232502 (2005).

[24] T. Terasawa, Progr. Theor. Phys. **23**, 87 (1960).

[25] T. Myo, A. Umeya, H. Toki and K. Ikeda, Phys. Rev. C **84**, 034315 (2011).

[26] T. Myo, A. Umeya, H. Toki and K. Ikeda, Phys. Rev. C **86**, 024318 (2012).

[27] H. Feshbach, Annals of Phys. **19** 287 (1962).

[28] A. Bohr and B. R. Mottelson, "Nuclear Structure", World Scientific Vol. I, p239.

[29] I. Hamamoto and S. Shimoura, J. of Phys. G **34**, 2715 (2007).

[30] B.A. Brown and B.H. Wildenthal, Ann. Rev. Nucl. Part. Sci. 38, 29 (1988).